\documentclass[useAMS,usenatbib]{mn2e}

\usepackage{psfig}

\title[2S\,1803$-$245 in quiescence]{An XMM-Newton observation of the neutron star X-ray transient 2S\,1803$-$245 in quiescence}
\author[R. Cornelisse et~al.]{R. Cornelisse$^{1,2}$\thanks{E-mail:
corneli@iac.es}, Wijnands, R.$^{3}$, Homan, J.$^{4}$\\
$^{1}$Instituto de Astrofisica de Canarias, Via Lactea, La Laguna E-38200, Santa Cruz de Tenerife, Spain\\
$^{2}$School of Physics and Astronomy, University of Southampton, Highfield, Southampton SO17 1BJ, UK\\
$^{3}$Astronomical institute ``Anton Pannekoek'', University of Amsterdam, Kruislaan 403, 1098 SJ, the Netherlands\\
$^{4}$MIT Kavli Institute for Astrophysics and Space Research, 77 Massachusetts Avenue, Cambridge, MA 02139, USA\\
}

\begin{document}

\date{Accepted  Received ; in original form }

\pagerange{\pageref{firstpage}--\pageref{lastpage}} \pubyear{2004}

\maketitle

\label{firstpage}

\begin{abstract}
  We observed the neutron star X-ray transient 2S\,1803$-$245 in
  quiescence with the X-ray satellite XMM-Newton, but did not detect
  it. An analysis of the X-ray bursts observed during the 1998 outburst of
  2S\,1803$-$245 gives an upper-limit to the distance of $\le$7.3 kpc,
  leading to an upper-limit on the quiescent 0.5-10 keV X-ray
  luminosity of $\le$2.8$\times$10$^{32}$ erg s$^{-1}$ (3$\sigma$).
  Since the expected orbital period of 2S\,1803$-$245 is several hrs,
  this limit is not much higher than those observed for the quiescent
  black hole transients with similar orbital periods.
\end{abstract}

\begin{keywords}
accretion, accretion disks -- stars: neutron -- stars:individual: 2S\,1803$-$245 -- X-rays:binaries.
\end{keywords}

\section{Introduction}

Low mass X-ray binaries (LMXBs) are compact binaries in which the
primary is a compact object (a black hole or neutron star) that
accretes matter from a low mass ($\le$1$M_\odot$) secondary. An
important sub-class of the LMXBs are the soft X-ray transients. These
systems spend most of their time in a quiescent state in which little
(or no) accretion is thought to take place and X-ray luminosities are
$\le$10$^{34}$ erg s$^{-1}$. Only occasionally do these transients
show outbursts that can last for weeks upto months and reaching X-ray
luminosities of 10$^{36-38}$ erg s$^{-1}$.

Over a dozen neutron star transients have been observed when they were
in the quiescent state, and in many cases their spectra show a soft
thermal component that is dominant below $\simeq$1 keV. This is
thought to be due to the cooling of the neutron star that has been
heated during the previous outbursts (e.g. Verbunt et~al. 1994; Brown
et~al. 1998; Campana et~al. 1998). Other mechanisms have also been
suggested to explain the quiescent emission for neutron stars, such as
residual accretion onto the neutron star (e.g. Campana et~al. 1998;
Campana \& Stella et~al. 2000). 

Apart from a soft component also a hard power-law component that
dominates the spectrum above a few keV has been observed in several
systems. This component can contribute a significant fraction to the
total X-ray flux, and especially in SAX\,J1808.4$-$3658 and
EXO\,1745$-$248 this power-law component was the main source of the
X-ray flux with no significant contribution from the soft thermal
component (Campana et~al. 2002; Heinke et~al. 2007; Wijnands et~al.
2005). The origin of this power-law component is still unclear,
although at low luminosities ($L_X$$<$10$^{33}$ erg s$^{-1}$) there
appears to be an anti-correlation between the fractional power-law
contribution to the luminosity and the source luminosity (e.g. Jonker
et~al. 2004).

One of the distinct differences between black hole transients and
neutron star transients is the difference in quiescent luminosities,
with the black hole transients being systematically fainter (e.g.
Narayan et~al.  1997; Menou et~al. 1999; Garcia et~al. 2001; Lasota
2007). This has been interpreted as evidence for the presence of an
event horizon in black holes. Since the energy is radiated away very
inefficiently for such very low accretion rates, this will not happen
before the matter has crossed the event horizon for black holes and
can therefore not be observed, while in neutron stars this should be
emitted at the moment the matter falls on the surface and can be
detected (e.g. Narayan et~al. 1997). However, alternative explanations
for this difference in luminosity have been suggested, such as a
transition to a jet-dominated regime for black hole transients that
carries away most of the material that would otherwise be accreted
(Fender et~al.  2003).

\begin{figure}
\psfig{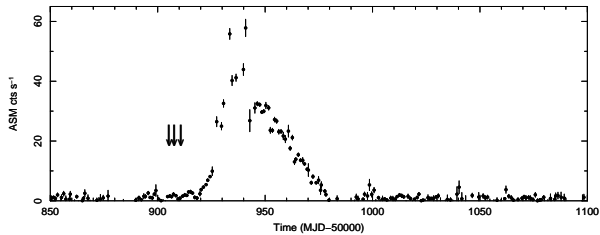}
\caption{Lightcurve of the outburst of 2S\,1803$-$245 obtained with the
All Sky Monitor. Indicated with arrows are the time of the 3 X-ray bursts 
observed with the WFCs.
\label{asm}}
\end{figure}

2S\,1803$-$245 (=XTE\,J1806$-$246) is a neutron star transient that
was first detected with the SAS-3 satellite in 1976 at a maximum
intensity of $\simeq$1 Crab (Jernigan 1976), and again during a second
outburst in 1998 that lasted for $\simeq$3 months (see
Fig.\,\ref{asm}) and that also reached a peak intensity of $\simeq$1
Crab (Marshall et~al.  1998). At the beginning of the second outburst
the BeppoSAX satellite detected thermonuclear X-ray bursts from this
source, establishing its neutron star nature (Muller et~al. 1998). A
radio counterpart was detected (0.8 mJy) that provided an accurate
position ($\alpha$=18h06m50.72s $\delta$=-24$^\circ$35$'$28.6$"$
J2000) and optical follow-up observations showed a weak (V$\simeq$22)
counterpart at the position of the radio source (Hjellming et~al.
1998; Hynes et~al.  1998).  During the peak of the outburst
2S\,1803$-$245 showed some spectral and timing properties of the
Z-sources (Wijnands \& van der Klis 1999), suggesting that it reached
accretion rates comparable to the Eddington rate (Hasinger \& van der
Klis 1989). Other observations during the decay of the outburst showed
that 2S\,1803$-$245 had the spectral and timing characteristics of
Atoll sources.

In this paper we report on an XMM-Newton observation of 2S\,1803$-$245
in quiescence, made $\simeq$7 years after its last outburst. Thus far, most
of the neutron star transients that have been studied in quiescence
showed sub-Eddington outbursts, with 2S\,1803$-$245 being one of the
few that reached the Eddington limit. This makes it an interesting
source to study how its quiescent properties compare to the other
systems.  However, in Sect.\,2.1 we will show that 2S\,1803$-$245 was
not detected during our observation. Combined with the analysis of the
X-ray bursts that were detected during its outburst (Sect.\,2.2) we
determine an upper-limit to the distance, and thereby an upper-limit
to its luminosity. In Sect.\,3 we will discuss the implications of our
findings.

\section{Observations and Data Reduction}

\subsection{Quiescent observations}

We made a 24 ks observation on 2S\,1803$-$245
using the X-ray satellite XMM-Newton from April 5 2005 (UT 22:23:52)
until April 6 2005 (UT 05:04:07). We analysed the data from the three
EPIC cameras (PN, MOS1, MOS2) that were observing in full window mode
and with a thin filter. The data were processed using the Standard
Analysis Software (SAS) version 7.0.0. In order to identify periods of
high particle background we extracted high energy ($\ge$10 keV)
lightcurves for all cameras. We chose to keep all data where the
countrate was less than 0.8 counts s$^{-1}$ for the PN and 0.2 counts
s$^{-1}$ for the MOS. This left a net observing time of 14.8 ks for the
PN and 20.3 ks for the MOS cameras.

\begin{figure}
\psfig{figure=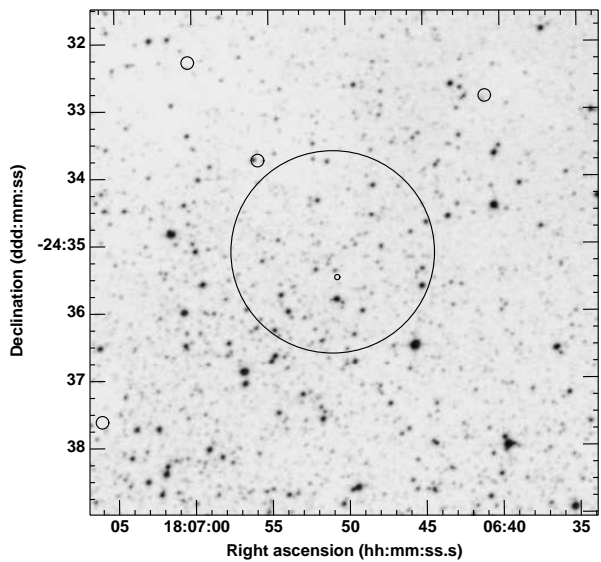,angle=0,width=8.cm}
\caption{Optical image from the Digitized Sky Survey of the region 
around 2S\,1803$-$245. The large circle is the RXTE error-circle, 
the small circle in the centre indicates the position of the radio 
counterpart of 2S\,1803$-$245, and the other circles indicate the 
positions of X-ray sources detected during the XMM-Newton observation. 
Note that the 4 arcmin error-circle from the BeppoSAX WFC is larger 
than the image.
\label{optical}}
\end{figure}

We created images for each individual camera for several energy ranges
(0.5-10, 0.5-2, 5-10 keV) but there was no detection of 2S\,1803$-$245
at the position of the radio source detected by Hjellming et~al.
(1998). Although we think it is unlikely, since its radio flux
(0.8$\pm$0.3 mJy at 4.86 Ghz) is similar to that of the bright neutron
star X-ray transients Aql\,X-1 and XTE\,1701$-$462 (both 0.5 mJy at
4.8 GHz; Fender \& Kuulkers 2001, Fender et~al. 2006), it cannot be
completely ruled that the radio source is not related to
2S\,1803$-$245. We therefore checked the region inside the RXTE
error-circle, but no source was present. In order to increase
sensitivity we also merged all 3 cameras and again created images in
different energy ranges. Still no source is present at the position of
the radio source, or even inside the RXTE error-circle (see
Fig.\,\ref{optical}). We therefore conclude that we have not detected
2S\,1803$-$245 in quiescence.

In order to determine an upper-limit on the X-ray flux of
2S\,1803$-$245 we extracted a spectrum for all 3 cameras using a
circle with a radius of 20 arcsec around the position of the radio
source. This lead to spectra with 9 counts for the PN and 4 counts for
each MOS detector.  For different spectral models, using the
absorption column determined in Sect.\,2.2 and combining all cameras,
we estimated a 3$\sigma$ upper-limit to the 0.5-10 keV unabsorbed X-ray
flux in Table\,\ref{upperlimits}. We have also compared these limits
with a source located closest to the radio position (see
Fig.\,\ref{optical}). This source was detected at 3.8$\sigma$ above
the background, and using the same spectral models as in
Table\,\ref{upperlimits} gave comparable flux levels as determined
for 2S\,1803$-$245. This makes us confident that the upper-limit on
2S\,1803$-$245 is correct. Using the upper-limit to the distance
determined from the X-ray bursts we also show the corresponding
luminosity in Table\,\ref{upperlimits}.

\subsection{Distance estimate}

2S\,1803$-$245 was in the field of view of the Wide Field Cameras
(WFCs; Jager et~al. 1997) onboard the BeppoSAX satellite (Boella
et~al. 1997) during its campaigns on the Galactic centre region.
During the campaign in the first half of 1998 three X-ray bursts were
detected from a position coincident with 2S\,1803$-$245.  Using the
publicly available data from the All Sky Monitor (ASM) onboard the
RXTE satellite we created a lightcurve of the outburst of
2S\,1803$-$245. In Fig.\,\ref{asm} we show its outburst, and have also
indicated the time that the bursts observed by the WFCs occurred.  We
note that all bursts occurred during the beginning of the outburst,
and assuming that the peak of the outburst was at the Eddington-limit
the X-ray luminosity must have been $\simeq$10$^{37}$ erg s$^{-1}$
(see below).  Since X-ray bursts are most commonly observed when a
source is at X-ray luminosities between 0.5-2$\times$10$^{37}$ erg
s$^{-1}$, but tend to be suppressed at higher luminosities (e.g.
Cornelisse et~al.  2003), we can be confident that they originated
from 2S\,1803$-$245.

\begin{figure*}
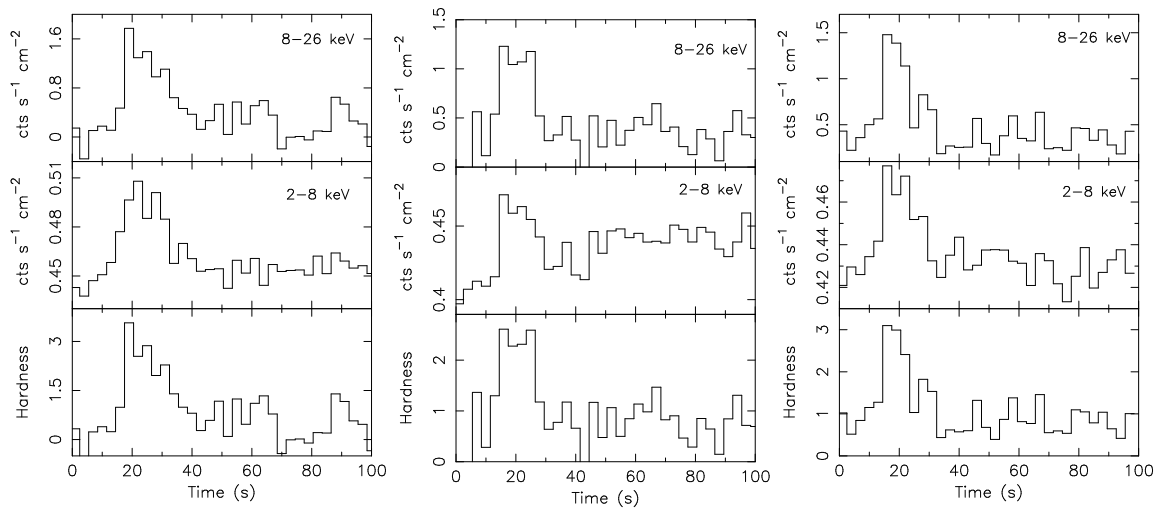

\parbox{5.8cm}{\psfig{figure=burst1.ps,angle=-90,width=5cm}}
\parbox{5.8cm}{\psfig{figure=burst2.ps,angle=-90,width=5cm}}
\parbox{5.8cm}{\psfig{figure=burst3.ps,angle=-90,width=5cm}}
\caption{Lightcurves (top and middle panels) and hardness curves 
(bottom panel) of the 3 X-ray bursts from 2S\,1803$-$245 observed with the 
BeppoSAX WFCs during its outburst. The times of occurrence of the 
X-ray bursts are MJD 50905.27136, 50907.67591 and 50910.76875 
respectively. For all curves the bin-time is 3s.  
\label{bursts}}
\end{figure*}

The X-ray bursts occurred between April 2 and 10 1998, and in
Fig.\,\ref{bursts} we show their 3 lightcurves in two different
energy-bands. The shapes of the bursts can be described by a fast rise
and exponential decay (with e-folding times between 10.1 and 13.5 s), as
is characteristic of a thermonuclear X-ray burst. Furthermore, we
have also calculated the hardness ratio (8-26 keV/2-8 keV) of the
bursts to show that spectral softening occurs during the burst.
Finally, we created a spectrum of the peak of the first (and
brightest) burst in order to estimate the corresponding flux.  The
spectrum could be well described by an absorbed black-body with a
temperature of 2.6$\pm$0.4 keV (and taking the absorption column fixed
at the value determined below), as is typically observed for
thermonuclear X-ray bursts. This translates into an unabsorbed
bolometric peak flux of 3.1$\pm$0.7$\times$10$^{-8}$ erg cm$^{-2}$
s$^{-1}$.

\begin{table}
\caption{3$\sigma$ upper-limits to the 0.5-10 keV unabsorbed X-ray 
flux of 2S\,1803$-$245 in quiescence for different spectral models. The
temperature (for the black-body model) and photon index, $\gamma$, (for 
the power-law model) are fixed at the indicated values, while for all 
models the absorption column is fixed at 1.47$\times$10$^{22}$ cm$^{-2}$.
Furthermore, we have indicated the corresponding luminosity for a distance
 of 7.3 kpc.
\label{upperlimits}}
\begin{tabular}{lccc}
\hline
Spectral Model & parameter & $F_{0.5-10}$ & $L_{0.5-10}$\\
               &           & (erg cm$^{-2}$ s$^{-1}$) & (erg s$^{-1}$)\\
\hline
black-body & kT=0.2 keV  & $<$4.4$\times$10$^{-14}$ & $<$2.8$\times$10$^{32}$\\
black-body & kT=0.5 keV  & $<$1.5$\times$10$^{-14}$ & $<$0.96$\times$10$^{32}$\\
Powerlaw & $\gamma$=1.5 & $<$2.0$\times$10$^{-14}$ & $<$1.3$\times$10$^{32}$\\ 
Powerlaw & $\gamma$=2.0 & $<$2.1$\times$10$^{-14}$ & $<$1.3$\times$10$^{32}$\\
\hline
\end{tabular}
\end{table}

Since the timing properties of 2S\,1803$-$245 suggested Z-source like
behaviour, and hence imply near-Eddington luminosities (Wijnands \&
van der Klis 1999), its persistent flux during the peak of the
outburst should be close to the peak flux of the X-ray bursts.
Although the quality of the data is not good enough to determine if
the X-ray bursts show radius-expansion, a clear indication that they
reached the Eddington-limit, we can still test if the persistent flux 
reached a similar level. From the log by Wijnands \& van der Klis
(1999) we selected the observation on May 3 1998 with the Proportional
Counter Array onboard the RXTE satellite (Jahoda et~al. 1996). This
observation showed the highest count rate, and also corresponds more
or less with the peak of the outburst according the ASM lightcurve in
Fig.\,\ref{asm}. The spectrum could be well fitted by a combination of
an absorbed black-body and absorbed disc black body models, with an
absorption column of 1.47$\times$10$^{22}$ cm$^{-2}$.  The unabsorbed
0.5-50 keV flux corresponds to 2.5$\times$10$^{-8}$ erg cm$^{-2}$
s$^{-1}$, which is close to peak flux of the X-ray burst, suggesting
that the outburst reached luminosities very close to the Eddington
limit.

\section{Discussion}

We have observed 2S\,1803$-$245 during its outburst in 1998 with
BeppoSAX and again in a $\simeq$20 ks observation in order to
determine its quiescent properties. We did not detect the source
during the XMM-Newton observations, and were only able to determine an
upper-limit on its quiescent flux of $<$4.4$\times$10$^{-14}$ erg
cm$^{-2}$ s$^{-1}$ (3$\sigma$). However, to compare this with other
neutron star transients in quiescence and the different cooling models
for neutron stars, we first need to determine the luminosity and
time-averaged mass transfer of 2S\,1803$-$245.

In order to determine its luminosity we presented the analysis of the
three X-ray bursts that were observed during the outburst of
2S\,1803$-$245. Since their peak flux was comparable to the continuum
flux during the peak of the outburst we can assume that they reached
the Eddington limit, which allows us to determine an upper-limit on
the distance. However, we must make several assumptions on the neutron
star properties in order to determine its Eddington limit. Since all
X-ray bursts showed an e-folding time of $\ge$10 s, indicative for the
presence of hydrogen during the burst (e.g. Fujimoto et~al. 1981;
Cornelisse et~al.  2003), we assume that 2S\,1803$-$245 has solar
metallicity.  Note that we can therefore not use the emperical
determined value of 3.8$\times$10$^{38}$ erg s$^{-1}$ by Kuulkers
et~al. (2003), since this is only valid for hydrogen-poor material.
Instead, we assume the canonical properties for the neutron star
parameters (i.e. radius of 10 km, mass of 1.4$M\odot$), leading to an
Eddington limit of 2$\times$10$^{38}$ erg s$^{-1}$. This leads to a
maximum distance of 7.3$\pm$0.7 kpc for 2S\,1803$-$245. Although the
formal error on the distance is only 10\%, due to the uncertainties in
the Eddington limit it will be larger. The largest uncertainty, as
suggested by the Eddington value determined by Kuulkers et~al.
(2003), is that the actual Eddington luminosity could be $\simeq$2
times larger than we used, leading to a distance that is at most
$\simeq$1.5 times larger than we estimated. The other uncertainty is
that the bursts do not show a clear indication of radius-expansion,
suggesting that they did not reach the Eddington limit. However, this
suggests that the Eddington flux for 2S\,1803$-$245 must be higher,
and therefore its distance lower than the upper-limit we determined
above. Despite these uncertainties we have used the distance value of
7.3 kpc to determine the upper-limit on the 0.5-10 keV luminosity
given in Table\,\ref{upperlimits}.

Following Tomsick et~al. (2004) we can estimate the
time-averaged mass transfer rate for 2S\,1803$-$245, $\dot M$, by
assuming that $\dot M$$=$$s$$L_{\rm peak}$$N$. Here $L_{peak}$ is the
peak luminosity, $N$ is the number of outbursts and
$s$$=$1.1$\times$10$^{-23}$ s$^2$ cm$^{-2}$ symbolising a value to
estimate the average accretion rate over a period of 33 years for a
source that has a similar outburst profile and duration as
XTE\,J2123$-$058 (see Tomsick et~al. 2004 for its outburst
lightcurve).  Since the outburst duration and the profile of
2S\,1803$-$245 is very similar to that of XTE\,J2123$-$058 we can use
this value of $s$. Given that 2S\,1803$-$245 has at least 2 outburst
over the last 33 years, and that it reached the Eddington luminosity,
we estimate an average mass accretion rate of $\dot
M$$=$7$\times$10$^{-11}$ $M_\odot$ yr$^{-1}$. Obviously, there are
many uncertainties in this value. For example, it assumes that we have
observed all outbursts of 2S\,1803$-$245 that occurred in the last 33
years, that all these outbursts were similar, that these 33 years
reflects the real time-averaged mass transfer rate. However, since it
is comparable to other estimates for the mass transfer rate, such as
using the time interval of the ASM lightcurve as done by Heinke et~al.
(2007), we think this value is currently the best we can derive.

We can compare the quiescent luminosity and average mass transfer rate
of 2S\,1803$-$245 with the predictions of the different cooling
models. Heinke et~al. (2007) did this for most other neutron star
transients that have been observed in quiescence (their Fig.\,2). As
has already been observed for many other systems (for overviews see
e.g. Cackett et~al. 2006, Heinke et~al. 2007), the quiescent
luminosity is too low to be explained by standard cooling models for a
low-mass neutron star as calculated by Yakovlev \& Pethick (2004).
This model predicts a luminosity that is at least an order of magnitude
higher than the upper-limit determined for 2S\,1803$-$245.
Only the models for more massive neutron stars, where the central
density is high enough to have more rapid direct Urca or Urca-like
processes, are consistent with our observations. However, we must note
that increasing the neutron star mass does increase its
Eddington-limit and thereby our estimate for the distance and
consequently increases both the upper-limit on the quiescent
luminosity and average mass transfer rate. Therefore, we cannot rule
out any of the other cooling models at the moment.

Although 2S\,1803$-$245 is fainter than expected for standard cooling
models, it is still an order of magnitude brighter than the currently
faintest neutron star transient 1H\,1905$+$000 (Jonker et~al. 2006).
At an upper-limit of 1.8$\times$10$^{31}$ erg s$^{-1}$ the luminosity
of 1H\,1905$+$000 is rivalling that of black hole transients in
quiescence (Jonker et~al. 2006). This system could challenge the idea
that black hole systems should have lower luminosities than neutron
star systems in quiescence (e.g. Narayan et~al. 1997).  However, Menou
et~al. (1999) predicted that this should only be the case for systems
with a similar orbital period. Since there is a strong indication that
1H\,1905$+$000 is an ultra-compact binary (Jonker et~al. 2006), it
should be able to reach luminosities lower than the average block hole
system (but not as low as a black hole transient with a similar
period). The orbital period of 2S\,1803$-$245 is currently unknown,
but Lasota (2007) gives a relation between the maximum outburst
luminosity and orbital period for an hydrogen dominated disk (his
formula 3). Using the maximum observed X-ray luminosity for
2S\,1803$-$245, we found that this would result in an orbital period
of 9 hrs. Although this is only a rough estimate, it strongly
indicates that 2S\,1803$-$245 is not an ultra-compact object.
Comparing the quiescent luminosity of 2S\,1803$-$245 with neutron star
and black hole transients which have orbital periods around 9 hrs (see
Garcia et~al. 2001), we note that it is located at the bottom of the
region where the neutron stars are located. More interestingly, the
current upper-limit is not that much higher than the luminosity of the
black holes. This makes 2S\,1803$-$245 an excellent candidate for deep
observations with the Chandra telescope to determine its quiescent
flux, and find out if it reaches X-ray luminosities comparable to the
black hole transients.

\section*{Acknowledgements}
We acknowledge Jean in 't Zand for providing the BeppoSAX Wide Field
Cameras data. We would like to thank the RXTE/ASM teams at MIT and
GSFC for provision of the on-line ASM data. We acknowledge the use of
the Digitized Sky Survey produced at the Space Telescope Science
Institute under U.S. Government grant NAG W-2166. RC acknowledges
financial support from a European Union Marie Curie Intra-European
Fellowship (MEIFT-CT-2005-024685).

\bsp

\label{lastpage}

\end{document}